# Two-dimensional ReS$_2$: Solution to the Unresolved Queries on Its Structure and Inter-layer Coupling Leading to Potential Optical Applications

## *(Accepted in Physical Review Material)*

Janardhan Rao Gadde[+,1], Anasuya Karmakar[+,2], Tuhin Kumar Maji[+,3], Subhrajit Mukherjee[4], Rajath Alexander[5], Anjanashree M R Sharma[6], Sarthak Das[7], Anirban Mondal[8], Kinshuk Dasgupta[5], Akshay Naik[6], Kausik Majumdar[7], Ranjit Hawaldar[1], K V Adarsh[8], Samit Kumar Ray[4,9] and Debjani Karmakar[*,10]

[1] *Centre for Materials for Electronic Technology, Pune, India, 411008*

[2] *Indian Institute of Science Education and Research, Pune, India, 411008*

[3] *Department of Chemical Biological and Macromolecular Sciences, S.N. Bose National Centre for Basic Sciences, Salt Lake, Kolkata, India, 700106*

[4] *Department of Physics, Indian Institute of Technology Kharagpur, Kharagpur, India, 721302*

[5] *Advanced Carbon Materials Section, Materials Science Group, Bhabha Atomic Research Centre, Trombay, Mumbai, India, 400085*

[6] *Centre for Nano Science and Engineering, Indian Institute of Science, Bangalore, India, 560012*

[7] *Department of Electrical Communication Engineering, Indian Institute of Science, Bangalore, India, 560012*

[8] *Department of Physics, Indian Institute of Science Education and Research, Bhopal, India 462066*

[9] *Department of Condensed Matter Physics and Material Science, S.N. Bose National Centre for Basic Sciences, Salt Lake, Kolkata, India, 700106*

[10] *Technical Physics Division, Bhabha Atomic Research Centre, Trombay, Mumbai, India, 400085*

[+] *These authors have equal contribution to this work*

*Corresponding Author: Dr. Debjani Karmakar (debjan@barc.gov.in)*

**Abstract**

Over the last few years, ReS$_2$ has generated a myriad of unattended queries regarding its structure, the concomitant thickness dependent electronic properties and its apparently contrasting experimental optical response. In this work, with elaborate first-principles investigations, using density functional theory (DFT) and time-dependent DFT (TDDFT), we identify the structure of ReS$_2$, which is capable of reproducing and analyzing the layer-dependent optical response. The theoretical results are further validated by an in-depth structural, chemical, optical and optoelectronic analysis of the large-area ReS$_2$ thin films, grown by the chemical vapor deposition (CVD) process. Micro-Raman (MR), X-ray photoelectron spectroscopy (XPS), cross-sectional transmission electron microscopy (TEM) and energy-dispersive X-ray analysis (EDAX) have enabled the optimization of the uniform growth of the CVD films. The correlation between the optical and electronic properties was established by static photoluminescence (PL) and excited state transient absorption (TA) measurements. Sulfur vacancy-induced localized mid-gap states render a significantly long life-time of the excitons in these films. The ionic gel top-gated photo-detectors, fabricated from the as-prepared CVD films, exhibit a large photo-response of ~ 5 A/W and a remarkable detectivity of ~ $10^{11}$ Jones. The outcome of the present work will be useful in promoting the application of vertically grown large-area films in the field of optics and opto-electronics.

I.  **Introduction**

The transition metal dichalcogenide (TMDC) ReS$_2$ had initially instigated a lot of expectations in the fields of optics and opto-electronics because of the prior prediction of its layer-independent electronic properties and thereby inferred direct band-gap from bulk to monolayer[1]. Few-layered ReS$_2$ had shown a reasonable photoresponse, stimulating a panoply of applications in photodetectors [2-4], phototransistors [5], logic circuits, [6-8] photocatalysis [9,10] and so on. ReS$_2$ differs from the common TMDC systems by virtue of its strongly anisotropic polarization-dependent optical [11-16], excitonic [17-20] and thermal responses[2,21-23], originating from its lower intra-plane symmetries in the distorted 1T′ structural format.

The system has continued to remain a source of a multitude of controversies over the last few years, since its structure, as predicted by Wildervanck *et al.* and Murray *et al.*[24,25], does not corroborate the electronic properties derived from a series of optical experiments, performed mostly on two-dimensional (2D) exfoliated single crystals and thin films with varying layer thicknesses. Although the X-ray diffraction measurements of the halide-assisted high-temperature-grown bulk single-crystals by Ho *et al.* had indicated the structure defined by Wildervanck *et al.*, the structural confirmation of 2D low-temperature growth of ReS$_2$ is still under cogitation and altercation [26]. The initial studies on ReS$_2$ predict the additional electron in Re to be responsible for the strong intra-plane Re-Re metallic bonds. Such metallic bonding leads to the formation of Peirl's distorted Re4 parallelogram array along an in-plane axis. As a result, there is a consequential and significant increase of the intra-plane interactions, rendering the inter-plane couplings negligible [1]. These studies predict ReS$_2$ to be a direct band gap semiconductor irrespective of its layer thicknesses [1]. The subsequent experimental studies, however, strongly oppose these predictions about the structural and electronic properties, revealing strong layer-dependent optical [11,27] and vibrational [28] properties of ReS$_2$. An additional indication of lesser quantum yields of photoluminescence with lowering thicknesses [29] also contradicts the direct band-gap nature of mono- and few-layered ReS$_2$. The optical response divulges that starting from mono-layer up to seven layers, ReS$_2$ remains an indirect band-gap semiconductor [28]. A couple of recent angle resolved photoemission (ARPES) studies on bulk and few-layered ReS$_2$ have unfolded a significant delocalization of valence electrons over the inter-layer van der Waals gap, nullifying the prediction of negligible inter-layer interactions [30-32]. Starting from the structure predicted by Murray *et al.*, it has also been demonstrated that except bilayer $ReS_2$, all other layered $ReS_2$ systems possess an indirect band-gap [30]. Moreover, few layered $ReS_2$ exhibits a series of inter-layer shear and breathing modes at frequencies below 50 cm$^{-1}$ [33-35], strengthening the probability of a strong inter-layer coupling in $ReS_2$. Therefore, the understanding of the correlation between the structural, optical and electronic properties of $ReS_2$ does not have the requisite transparency in spite of the existence of numerous studies in prior literature.

An exhaustive survey of the prior results has indicated the existence of two possible structures of $ReS_2$ in the literature. The cell parameters, unit cell orientations and polyhedral Re-S coordination of these two structures are widely different. The structure defined by Murray *et al.*[24] possesses a distorted octahedral coordination and ignores one centre of reflection. Later on, Lamfers *et al.* [36] have suggested the correct structure of $ReS_2$, with two types or polyhedral coordination, modified cell parameters and unit-cell orientation, where they claimed to have rectified the errors in the structure of Murray *et al.* Interestingly, all prior theoretical studies have concentrated only on the structure by Murray *et al.* [24] and the structural rectification suggested by Lamfers *et al.* [36] remained apparently unattended.

In the present work, we aim at filling up the existing voids in the understanding of the electronic properties of $ReS_2$ with the help of theoretical analysis and subsequent experimental validations. The organization of the present work is as follows. In the next section, we have furnished a comparison between two different structural configurations of $ReS_2$, available in the literature and with a thorough first-principles investigation have identified the correct one, capable of reproducing the experimental results. Subsequently, synthesis of large-area thin film was carried out and a stoichiometric optimization of the CVD growth process was obtained with the help of spectroscopic characterizations, as described in the next section. As a next step, well-characterized as-grown large-area systems are utilized for demonstrating optical and opto-electronic applications. The last section provides a summary and conclusion of the obtained results.

II. Theoretical Analysis: Correlation between structure and properties

$ReS_2$ belongs to the triclinic symmetry group P$^1$, resembling distorted $CdCl_2$ structure. Intricate scrutiny of prior literatures regarding its structure leads to two different results. The structure defined by Murray *et al.*[24], having four formula units per unit cell, is rectified by Lamfers *et al.*[36], with a modified orientation of the unit cell, numerical values of the cell parameters and the underlying inversion symmetry. Throughout this work, we will define the structures described by Lamfers *et al*. and Murray *et al*. as S-1 and S-2 respectively. The S-1, possessing eight formula units per unit cell, has four almost coplanar independent crystallographic sites for Re due to the pseudo-inversion center at (1/4, 0, 0), the repetition of which along the *c*-axis is related by a weak reflection symmetry. This center of reflection was ignored at S-2, leading to a halving of its *c*-axis. In addition, for the orientation of the unit cell in S-2, the ReS$_2$ planes are stacked along the *a*-axis, in contrast to the *c*-axis stacking derived for S-1. A pictorial comparison of these two structures and their corresponding unit cell orientations is presented in Figure 1. As per our knowledge, maximal prior studies refer to the structure as in S-2 and the structure defined by S-1 remains disregarded till date. In the present work, we have compared the energetics and the electronic band properties corresponding to these two structures.

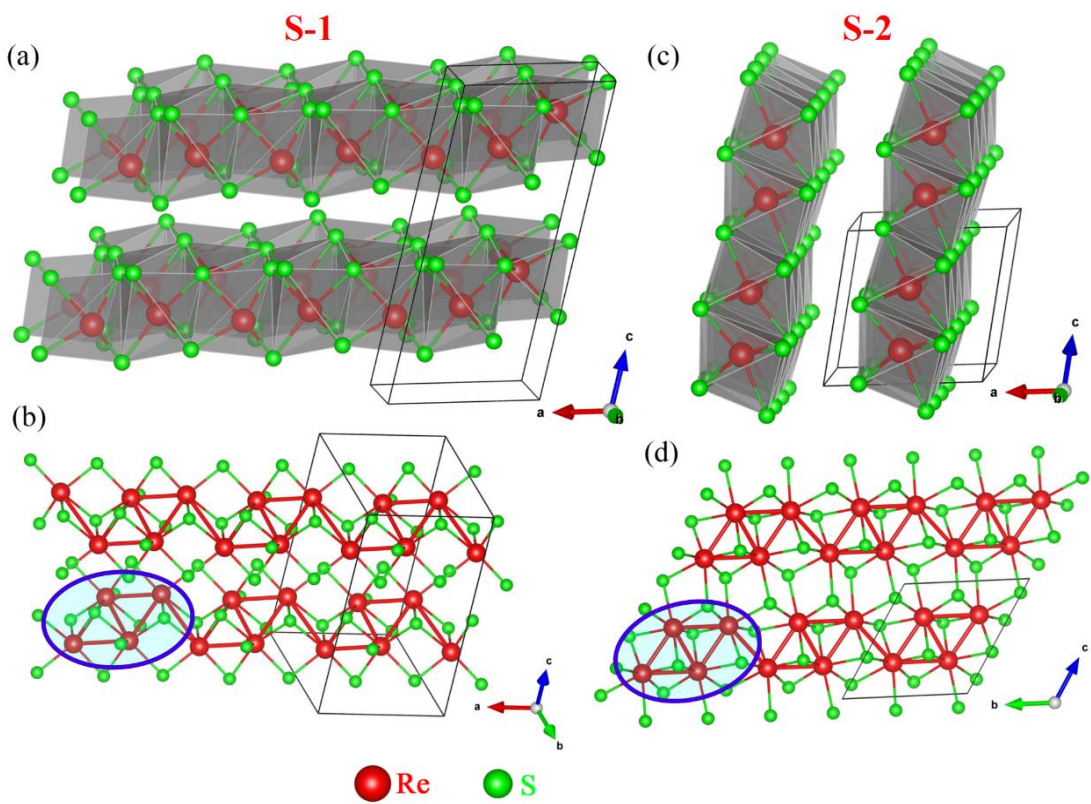

*Figure 1: Polyhedral coordination of OCT and SQP for a) S-1 with the corresponding unit cell orientation and (b) the Re4 parallelogram array along the a-direction (a typical Re-4 parallelogram highlighted within blue ellipse) ; (c) the OCT polyhedral coordination for S-2 with the respective unit cell orientation and (d) the Re4 parallelogram array along the b-direction (highlighted within blue ellipse). Red and green are the colors are used for indicating Re and S respectively.*

The main characteristic, which differentiates these two structures, is their respective polyhedral coordination. For S-1, there are two types of polyhedra. On *a-b* plane, the face-shared distorted ReS$_6$ octahedral (OCT) chains along *b*-axis are bridged by the two face-shared couplet of ReS$_5$ distorted square pyramids (SQP) along *a*-axis. Figure 1 (a) and (c) present a comparison of the polyhedral coordination and the corresponding unit cell orientations of S-1 and S-2 respectively. There is only one type of polyhedral coordination in S-2, where the face-shared OCT pervades the *b-c* plane. The Re4 parallelogram chains, as shown within the highlighted ellipse, are aligned along the *a* and *b* axis for S-1 and S-2 respectively (Figure 1 (b) and (d)).

Comparative analysis of the electronic structures of both S-1 and S-2 was carried out by first-principles calculations using the generalized gradient approximation (GGA) with a hybrid BLYP exchange correlation functional, after incorporating the spin-orbit coupling. The details of the parameters of the theoretical calculations are presented in the supporting information [37]. To compare the formation energies of S-1 and S-2, we have taken a 1×1×2 supercell of S-1 and a similar supercell of S-2, constructed from its *a-c* flipped unit cell. The formation energies of both of these structures, calculated by using the formula, $E_{form} = E(ReS_2)_n - nE(Re) - 2nE(S)$, are found to be negative, supporting the stability of both the structures. Nevertheless, the formation energy per unit cell of S-1 is ~100 meV below that of S-2, implying a better stability for S-1. However, a comparison of this energy difference with the computed exfoliation energies of layered ReS$_2$ system reveals an interesting property of this structure. The exfoliation energies for ReS$_2$ are computed by using the formula [38], $E_{ex}(n) = (E_{slab}(n) - E_{bulk}/m)/A$, where $E_{ex}(n)$ is the exfoliation energy of *n* layers, $E_{slab}(n)$ is the energy of the unit cell of the *n*-layer slab in vacuum, $E_{bulk}$ is the energy of unit cell of the bulk system consisting of *m*-layers and *A* is the in-plane area of the bulk unit-cell. The exfoliation energy of the monolayer of ReS$_2$ is ~14.7 meV/ Å$^2$ and it rises sharply for bi- and tri-layers to ~22 and 31 meV/ Å$^2$ respectively and soon reaches saturation from 5-layers onwards at ~52 meV/ Å$^2$. Such sharp rise of exfoliation energies explains the frequent use of ultrasonic or chemical exfoliation technique for experimentally producing few-layered ReS$_2$ flakes [38-40]. Thus, few layered ReS$_2$ (S-1) systems indicate comparatively larger exfoliation energies than other common 2D systems [38]. In addition, within an area of the order of Å$^2$, exfoliation of few-layered system may lead to a metastable cross-over between S-1 and S-2, as indicated by the difference of their respective formation energies.

As an obvious next step, we have calculated the layer-dependence of the electronic structure of mono-, bi-, tri-layer and bulk systems of S-1, where the band-dispersions and the corresponding partial density of states (PDOS) are plotted in Figure 2(a) – (d) respectively. The high-symmetry directions are shown with respect to the corresponding Brillouin zone in Figure 2(f). Intriguingly, the band-dispersions indicate an indirect to direct band-gap cross-over, with an indirect band gap of 1.46 eV for monolayer, which gradually reduces due to the energy-shift of the valence band maxima, resulting into a direct band gap of 1.32 eV for the bulk system. Fig 2(e) depicts the in-plane bridging of the face-shared distorted $ReS_6$ octahedral (OCT) chains along *b*-axis through the two face-shared couplet of $ReS_5$ distorted square pyramids (SQP) along *a*-axis.

Figure 2(g) and (h) presents the static and time-dependent optical absorbance per unit optical path calculated by using DFT and TDDFT respectively, which is compatible with the corresponding band-structures. In both of these cases, with increasing number of layers, there is an increase in the value of the absorbance. The calculated electronic structure and the corresponding optical properties match very well with the experimental results existing in the literature, where with increasing thickness of $ReS_2$ layers, there is an increase in the intensity of luminescence [29].

Presence of metallic clusters and the directionality of its in-plane orientation results in highly anisotropic optical response for $ReS_2$, as is also observed experimentally [11]. Figure S1 presents the static anisotropic optical absorbance per unit optical path from monolayer to bulk system, where all the three components of the absorbance are inexorably different.

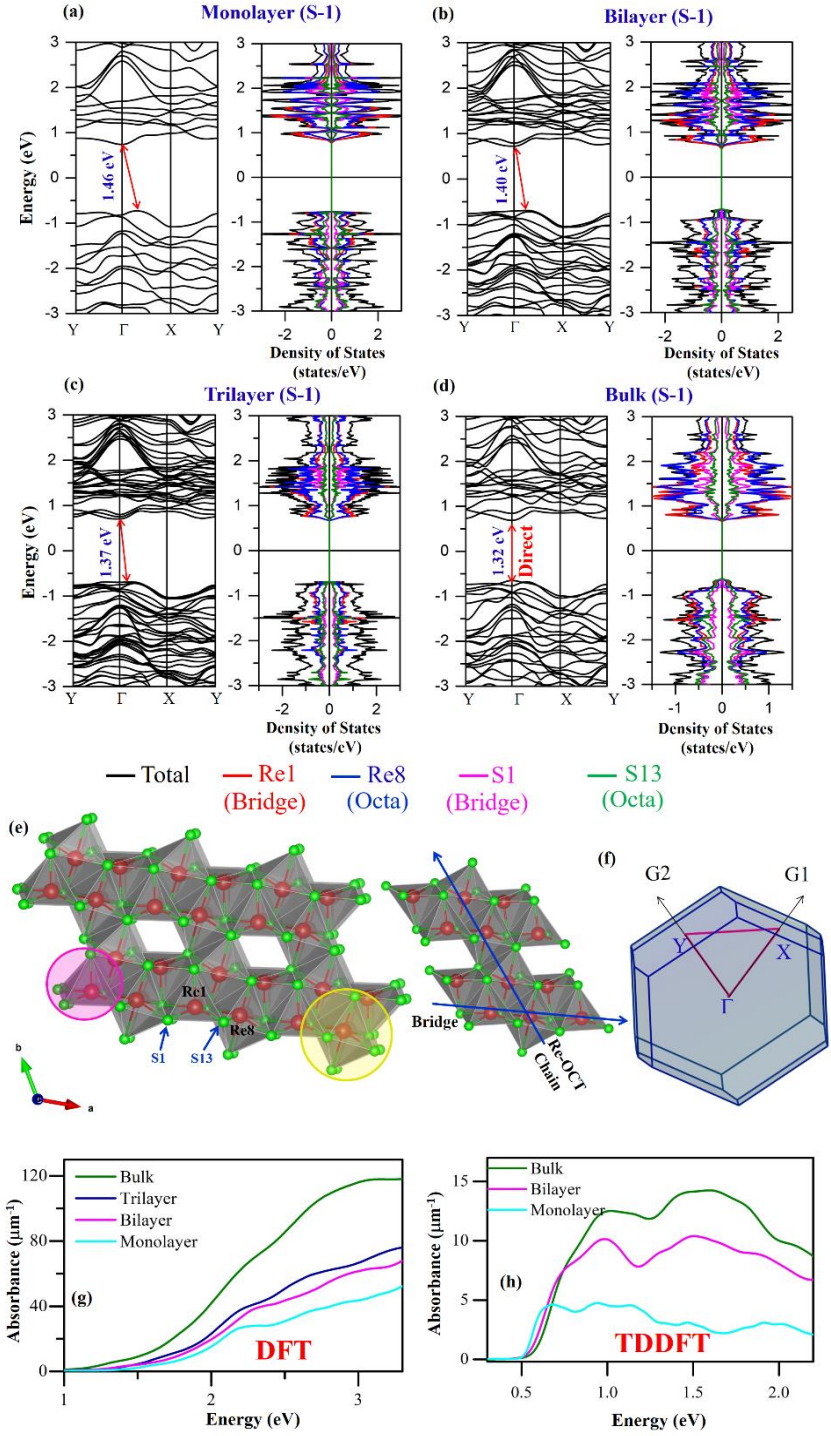

*Figure 2:* Band structure and corresponding DOS of ReS$_2$ (S-1) a) monolayer, b) bilayer, c) trilayer and d) bulk system. e) Multiple polyhedral coordination of ReS$_2$ in S-1, square pyramidal (SQP) and octahedral (OCT) polyhedra have been marked by using pink and yellow circle respectively. Right inset shows the 1D OCT chain and bridging SQP structure f) corresponding Brillouin Zone and the high symmetry directions of the structure. Optical absorbances of different ReS$_2$ systems are plotted after using g) DFT and h) TDDFT methodologies.

The band-dispersions for S-2 are plotted in Figure 3, which closely resembles the reference [30], where the system indicates an indirect band-gap all-along, except for its tri-layer with a direct band-gap of 1.28 eV.

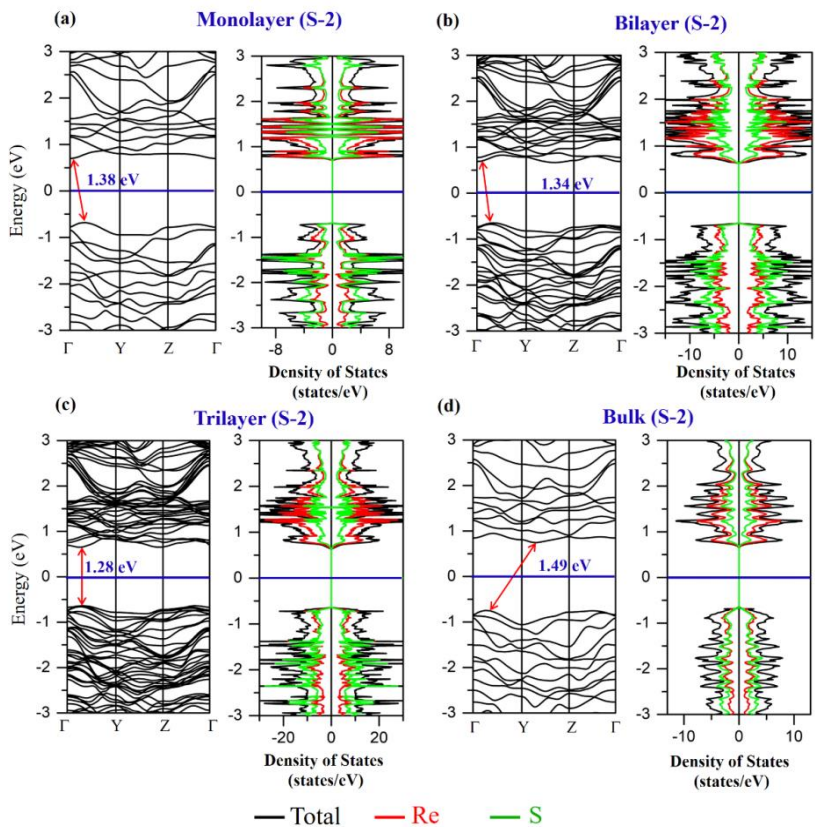

*Figure 3:* Layer dependence of band structure and corresponding PDOS for S-2 of ReS$_2$ in 1T- Octahedral coordination for a) Monolayer, b) Bilayer, c) Trilayer, and d) Bulk.

Intricate scrutiny of the structural details of S-1 reveals that the S-atoms belonging to OCT and SQP are in different planes and the inter-planer distances between the S atoms belonging to the different layers are as short as ~ 3 Å. Thus, S-1 promotes the possibility of inter-planer electronic cross-talking much stronger than the dipolar van der Waals interactions, leading to an increase in the exfoliation energies with the number of layers. Moreover, the distortion of the SQP bridging couplet and the strong hybridization of its center Re atom (Re1) with the constituent S-atoms, which are corner-shared with the OCT chain (S1 and S13, Figure 2(e)), have a prominent role on the layer-dependence of the electronic structure. The states at the top of the valence band and the bottom of the conduction band are mostly populated by the hybridized states of S13 and S1 with Re1 (Figure 2(a)-(d)) and are therefore partially responsible for the band-gap cross-overs. Additionally, there is a significant hybridization of the terminating S atoms of a layer with the center Re atoms of OCT and SQP of the next layer via their respective constituent edge S atoms, as will be evident from the Figure 4 (a) – (d) for bulk, mono, bi- and tri-layers of ReS$_2$. This hybridization becomes more prominent with increasing number of layers and may provide another reason behind the indirect to direct bandgap cross-over from monolayer to bulk. Thus, both intra- and inter-layer polyhedral Re-S hybridizations are culpable for the layer-dependent electronic response.

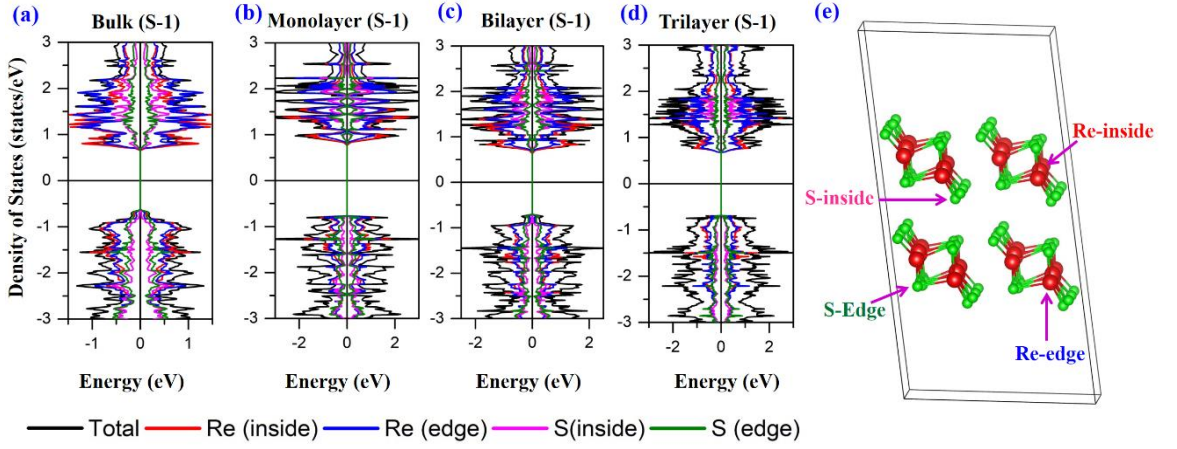

*Figure 4:* Contribution of edge and inside Re and S atoms in the density of states (DOS) of a) bulk, b) monolayer, c) bilayer and d) trilayer, Figure (e) denotes the edge and inside atoms taken into consideration in the PDOS.

ReS$_2$ is also well-known to have various possible defects [41], actuating its use in producing solid-state lasers [42]. Figure 5 depicts the role of S-vacancies at the (a) OCT, (b) SQP and (c) both sites, indicating the presence of highly localized mid-gap levels and an associated *n*-type doping, as seen in other TMDC too [43].

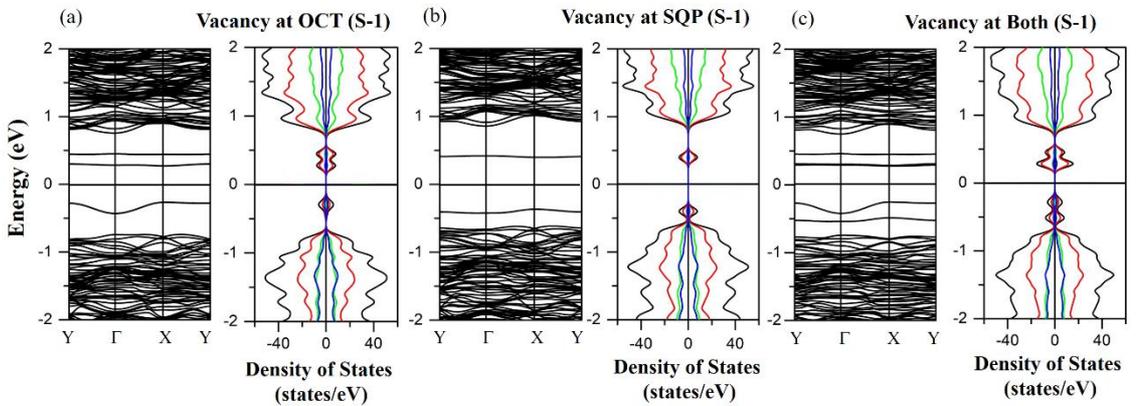

*Figure 5:* Band structure and corresponding PDOS of bulk ReS$_2$ (S-1) with Sulfur vacancies at the a) distorted octahedra (Vacancy at OCT), b) Square pyramid (Vacancy at SQP) and c) at both polyhedral places (Vacancy at Both).

In the next section, we have carried out the experimental investigations on large-area ReS$_2$ samples to endorse our theoretical results.

### III. Experimental analysis: Synthesis and Characterization

With a clearer idea about the structural details, we have looked for an experimental validation of the theoretically obtained results, initiating with the synthesis of ReS$_2$ in the proper phase. There are numerous effective methods to synthesize a single-phase multilayered ReS$_2$ system starting from the top-down non-scalable methods like mechanical or liquid exfoliation [39], high-temperature halogen-assisted growth of single crystals [44] to the scalable bottom-up epitaxial growth process like chemical vapor deposition (CVD) [45-47]. Synthesis of ReS$_2$ by direct sulfurization of ammonium perrhenate (NH$_4$ReO$_4$) is known to be an effective and relatively low-temperature process to obtain the large area films at an ambient pressure [46,48-50]. The details of this synthesis procedure are described in the supporting information [37]. In this work, single phase multilayered ReS$_2$ thin films are grown on the Si/SiO$_2$ wafers at three different growth temperatures, *viz.*, 450, 650 and 750 °C. We designate the films grown at these three different temperatures as F1, F2 and F3 respectively. Unlike the other TMDC systems, ReS$_2$ is well-known for its inherent spontaneity towards a non-planar vertical wall-like or ribbon-like CVD growth, irrespective of the choice of the substrate [48-53]. Several propositions attempt to unravel the reason behind this vertical mode of growth, including the mismatch between the rates of the supply of the precursors to their respective rates of diffusion and the stress induced curling, originating from the lattice mismatch with the substrate [51]. Ghosal *et al.* demonstrated the two-step nature of the CVD growth, *viz.*, the initial horizontal growth providing a full coverage to the substrate, followed by the successive vertical ones that originate from the out-of-plane protrusions of the octahedral S-atoms and the Re-Re metallic bonds [48]. Such vertical growths, although demonstrated for an improved applicability in catalysis, field emission, energy storage and water purification [48,51], are scarcely validated for their pertinence in the field of excited state photo-response and opto-electronics. In the current experimental part of this work, we have emphasized over such applications.

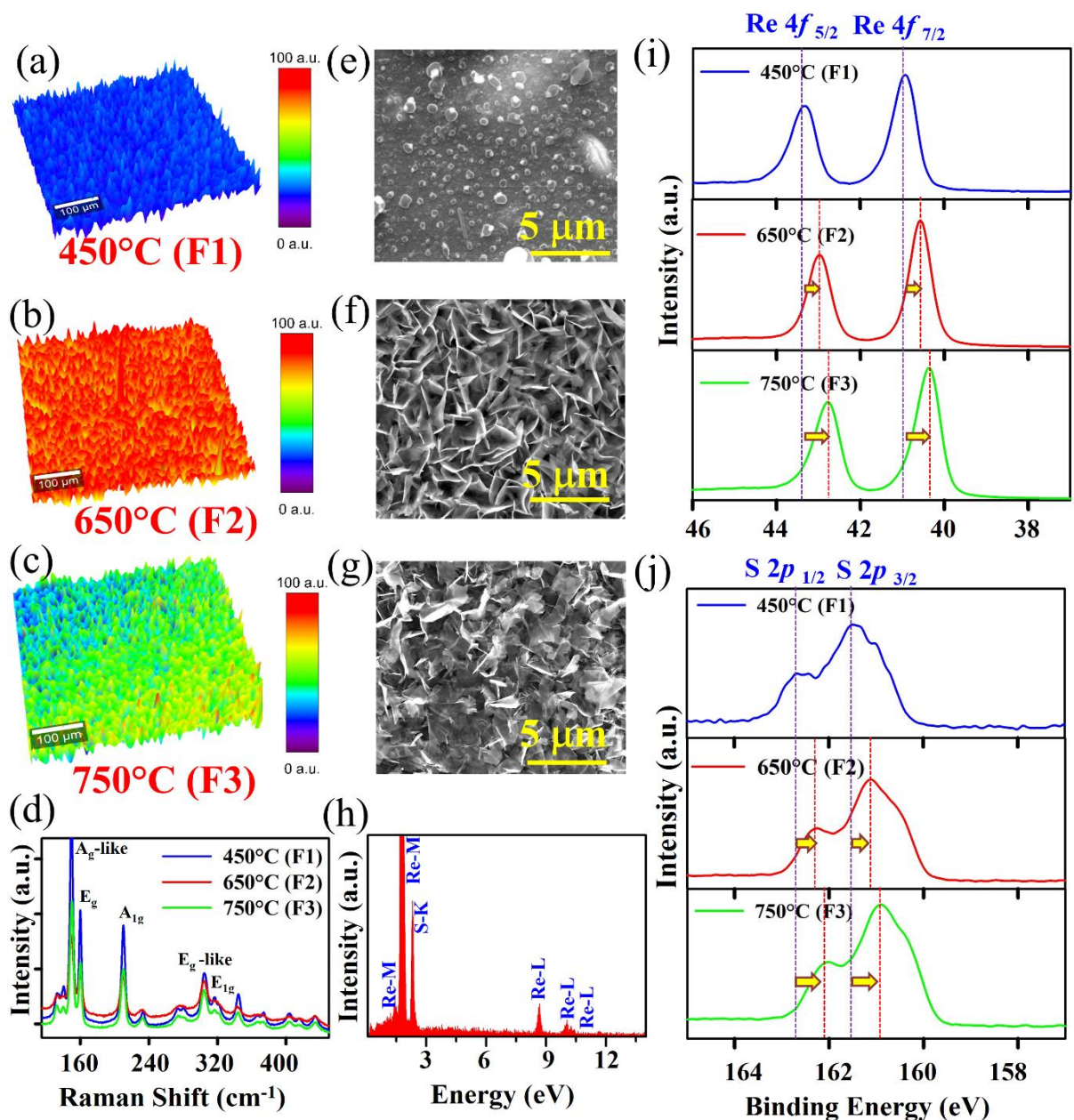

*Figure 6:* *3D Raman Maps of different CVD grown films at different growth temperatures: a) 450°C (F1), b) 650°C (F2), c) 750°C (F3) and their corresponding d) Raman Spectra. Corresponding SEM images of different films for e) 450°C (F1), f) 650°C (F2), g) 750°C (F3); h) EDX elemental map of $ReS_2$ (650°C); XPS Spectra of (i) Re-4f and (j) S-2p electrons in different samples (F1-F3).*

Figure 6 presents the micro-Raman (MR), energy dispersive X-ray (EDX) analysis and the X-ray photoelectron spectra (XPS) as a function of the different growth temperatures. As supporting evidence towards a uniform large area deposition of the proper crystalline phase, the MR spectra for the feasible Raman active modes were reproduced from the average of several Raman mappings over an area of 500 μm × 500 μm at various places of each film with a laser of wavelength 532 nm and a spot size of ~20 μm. For $ReS_2$, unlike the other TMDC systems, the lack of symmetry, intra-mode couplings and the interactions with acoustic modes incite 36 vibrational modes altogether. Keeping apart the 15 infra-red active and 3 zero-frequency modes, the rest 18 Raman active ones constitute the $A^g$-like and $E^g$-like vibrational modes, originating from the out-of-plane and in-plane vibrations of the Re-atoms respectively [54]. Figures 6(a) – (c) depict the Raman intensity maps with the associated color scales, after considering the most intense $A^g$-like (149 cm$^{-1}$) peak, having much less sensitivity towards polarization. The Raman maps evinced that F2 (650°C) is having the most uniform and intense Raman spectra, as is also manifested in the MR spectral plots for F1—F3 in Figure 6(d). The peak positions for the possible Raman modes, corresponding to the Figure 6(d), are presented in Table I for F1-F3, indicating negligible frequency shifts. The $A^g$ and $E^g$-like peaks occur at ~ 149 cm$^{-1}$ and 304 cm$^{-1}$ respectively. Higher energy $E^g$-like modes and other peaks originate from the out of plane vibrations of the S-atoms. In addition to those, two prominent modes $E^g$ and $A^{1g}$ are observed at 160 and 210 cm$^{-1}$ respectively [50,55]. Figure 6(e) – (g) represent the scanning electron microscopic (SEM) images for all the three films F1-F3. With increasing temperature, the two-dimensional film F1 (Figure 6(e)) evolves to obtain a uniformity for the wall-like growth at F2 (Figure 6(f)), which are disrupted with the increase in growth temperature, as in F3 (Figure 6(g)). The corresponding elemental map of F2 is presented in Figure 6(h) and Table I presents the percentage compositions of Re and S for F1-F3, with a clear indication of their 1:2 ratios for F2. Therefore, F2 turns out to be the most optimally grown film, in terms of the perfect stoichiometry of ReS$_2$. To obtain a correct compositional confirmation and the corresponding valence configuration, the binding energy shifts of the core level electrons are measured with high-resolution X-ray photoelectron spectroscopy (XPS) with hard X-rays in an energy-range of 2-15 KeV. The XPS spectra for the Re-4$f$ and S-2$p$ peaks are presented in Figure 6(i) and (j) respectively, which connotes the shift of these peaks towards the lower binding energy as a function of the growth temperature from F1 to F3. This shift for higher temperature growths may be associated with the presence of defects, inducing highly localized mid-gap states [56,57]. For the optimized sample F2, whereas the Re-4$f^{5/2}$ and 4$f^{7/2}$ peaks ensued at 42.99 and 40.55 eV, the S-2$p^{1/2}$ and 2$p^{3/2}$ peaks emanate at 160.3 and 161.1 eV respectively. The peak positions for F1-F3 are conferred in Table II and also compared with the single-crystal ReS$_2$ data [44]. The general observation for the CVD grown films points out the lower binding energies for the characteristic core-level peaks in comparison to the single-crystalline samples, as was also ascertained in earlier studies [44,46].

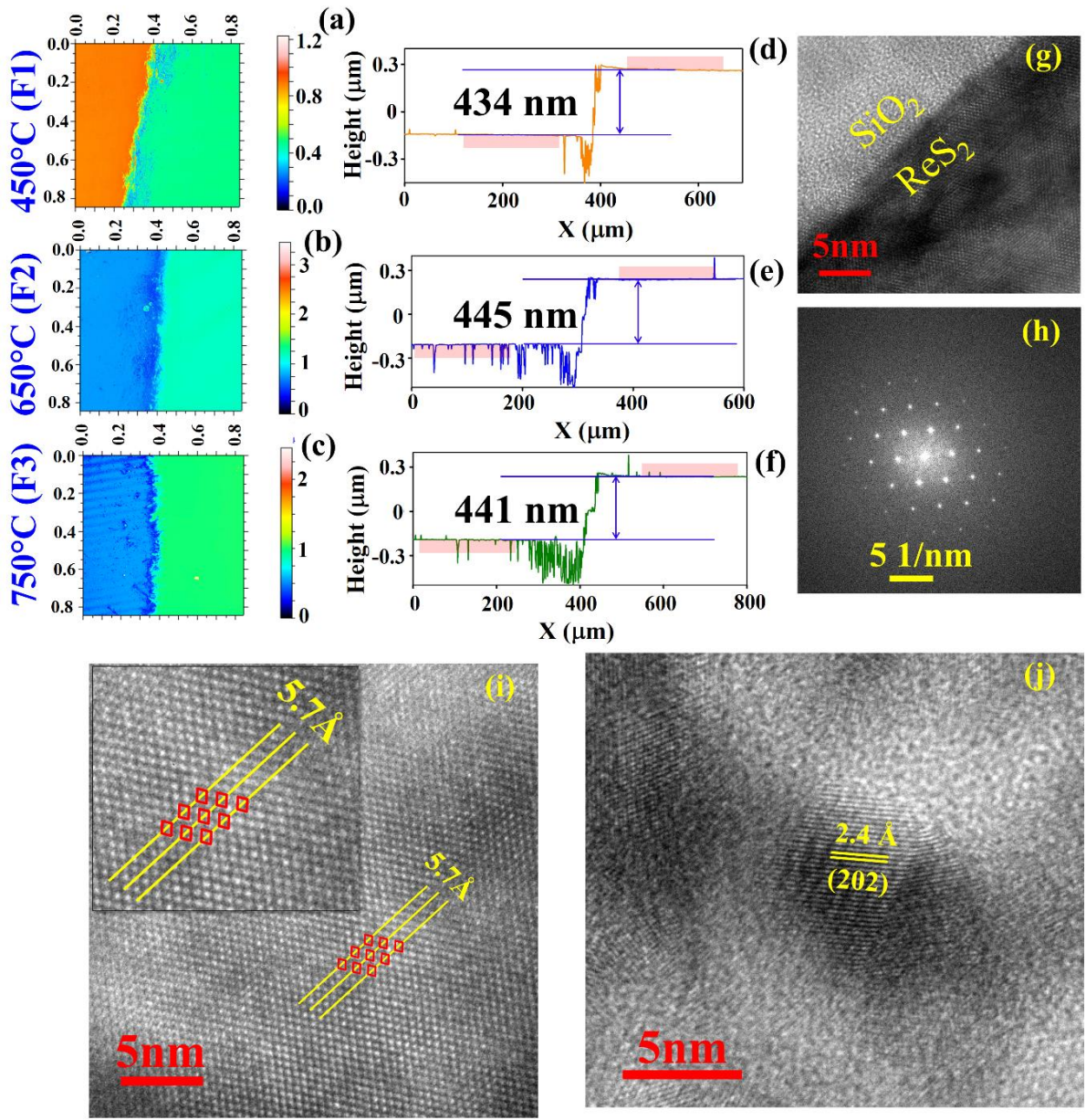

*Figure 7:* Colour map of thickness of ReS$_2$ samples using an optical profilometer for growth temperatures a) 450°C(F1), b) 650°C(F2) and c) 750°C(F3); Corresponding height profile of systems for d) 450°C(F1), e) 650°C(F2) and f) 750°C(F3); (g) cross-sectional TEM images of ReS$_2$ on SiO$_2$(F2), (g) FFT image of ReS$_2$, (i) HRTEM image of planar ReS$_2$ depicting the Re4 parallelogram, (j) HRTEM image of vertical growth of ReS$_2$ showing lattice fringes corresponding to the [202] plane of S-1(see text).

Figure 7(a) – (c) present the thickness profile of F1-F3, as measured after scanning the surfaces by an optical profilometer by using highly focused white light beam. The uniformity of the area-coverage in the samples is quite evident from the corresponding color plots. The corresponding height profiles are depicted in Figure 7(d)-(f) for all the three systems F1-F3. While analyzing the height profile, we emphasize on the two step growth processes of ReS$_2$. Upto 450°C, there is only horizontal growth, having a thickness of ~434 nm. With increase of growth temperature, the vertical growth initiates. The height profiles of sample F2 and F3, conveying the total thickness, may appear misleading due to their additional vertical growth on top of the horizontal film. Therefore, for all these three systems, the thickness of the horizontal crystallites covering the film can be taken as a few hundred nm, ascertaining them to be in the bulk regime. However, this thickness profile does not convey anything about the horizontal size of the crystallites. For more in-depth investigation of the atomic-scale structural properties, we have carried out the cross-sectional transmission electron microscopy on the $ReS_2$/$SiO_2$/Si interfaces for the sample F2 and the outcome is plotted in Figure 7(g)-(j). Figure 7(g) clearly represents the growth of the $ReS_2$ films on Si/$SiO_2$ substrate, having an uneven interfacial morphology due to the diffusion of $ReS_2$ within $SiO_2$. As is also observed in the prior studies [48], there are two different types of regions for these thin films, *viz.*, the top vertical growth and a flat coverage of $ReS_2$ beneath the vertical structure. Figure 7(h) depicts the fast Fourier transform (FFT) image, which clearly conveys the six-fold symmetry of the pattern, confirming the high crystalline nature of the films. When the electron beam is focused on the flat area, the HRTEM image clearly shows the Re4 parallelogram cluster arrays covering the entire region having an average distance of 5.7Å between the two arrays, as obtained from the figure 7(i). This distance closely resembles the same value as that obtained from the theoretical calculations (5.4 Å) for the structure S-1. The sides of the parallelogram range from 2.76 - 2.84 Å from the measurements, which are comparable with the theoretically obtained lengths of 2.80 - 2.85 Å. The corresponding distance between the parallelogram arrays for S-2 is 5.1 Å with the sides varying from 2.3 – 2.4 Å. Albeit the dimensions related to the Re4-parallelogram arrays are comparable for both structures, with S-1 being more close to the experiment, there are more details to be explored from the TEM micrograph. While the beam is focused on the vertical growth pattern, as in Figure 7(j), the obtained fringes imply an interlayer distance of 2.4 Å, corresponding to the [202] planes of the theoretically predicted structure S-1. Thus, the structural details, as obtained from the experimental large-area growths of $ReS_2$ indicate our experimental structure to be closer to S-1.

We may also mention in passing about two important aspects of the as-grown films. First, the theoretically obtained higher range of exfoliation energies for the structure S-1 was supported by the unsuccessful attempts to the mechanical exfoliation from the as-grown CVD films, leading to a peeling of the entire film from the substrate. Second, the XPS results imply that the as-grown films, especially grown at higher temperatures, are having large amount of defects in comparison to single crystals.

The structural analysis of the as-grown CVD films indicated the evidence of the presence of the structure S-1 at a local level. However, the complicated process of CVD growth, initiating in a planar geometry and followed by a cross-over to a vertical one, results into a complex composite film material. There is further need of more sophisticated and intricate analysis at both micro- and macroscopic level to unambiguously unravel the structure-properties correlations, which is beyond the scope of the present work.

In the next section, we have performed the static and dynamic optical measurements on these systems to highlight the applicability of these vertically grown systems in the field of optics and opto-electronics.

### IV.     Static and dynamic optical response of the CVD films

Both steady-state photoluminescence (PL) and excited state transient absorption measurements are performed on the samples F1-F3 to obtain an assessment about their potential optical attributes. PL was measured with a laser of wavelength 532 nm, a spot size of 20 μm diameter and the maximum laser power of 1.6 mW. The main feature of the PL spectra of $ReS_2$ is its significantly low quantum yields for the bulk samples F1-F3, resembling the prior studies. This outcome can be explained from our derived band-structure. The difference between the indirect band-gap for few-layered samples and the direct band-gap for bulk system is ~ 0.1 eV, as seen from Figure 2. Such small difference in the band-gap allowed the probabilities of the other transitions and thus broadens the peak and produces different shapes for the line-profiles. In Figure 8(a), For F1, there is presence of both A and B excitons with the intensity of the latter being higher, as is also seen in many prior studies [11]. The higher B exciton is attributed to the presence of a rapid non-radiative relaxation associated with the presence of Sulfur vacancy (SV) defects [11] or the lower oscillator strength of exciton A as compared to the exciton B. As can also be seen from the Figure 5, the SV defects create highly localized shallow traps within the band-gap of $ReS_2$, providing a suitable means for relaxation of the conduction electrons. For the other two samples F2 and F3, only a single exciton (B) signal is present. With increasing vertical growth, there will be a simultaneous increase in the defect densities, rendering a suppression of the A exciton. The PL peak of B exciton is at ~ 1.40 eV for F1 and with an increase in the growth temperature, there is a small blue shift in the peak positions, indicating stronger electronic couplings for high temperature samples. Additionally, the smaller exciton binding energy in $ReS_2$, as seen by Aslan *et al.*[11], can be reproduced from the difference of the theoretically calculated optical band-gap (1.32 eV) and the experimental exciton positions (1.40 eV). In addition to the lower energy exciton peaks, the samples display higher energy peaks from 1.8 - 2.0 eV, as depicted in Figure 8(b), the presence of which can be explained from the PDOS figure of the bulk system (Figure 2(d)). The VB and CB DOS peaks at ~ -1eV and 1 eV respectively, corresponding to the S-hybridized OCT and SQP Re-5$d$ states, are responsible for the origin of the higher energy peaks.

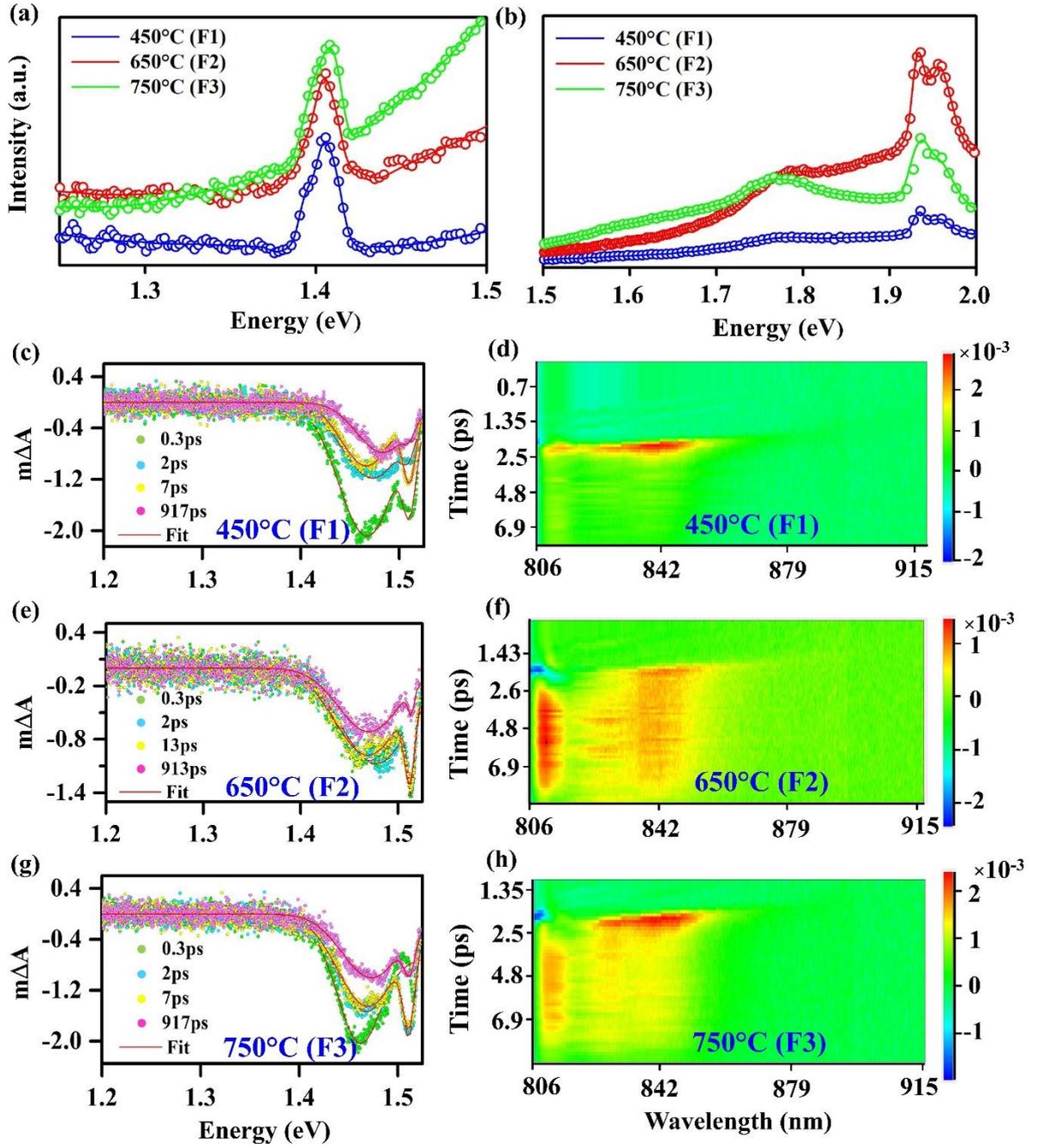

*Figure 8: Room temperature PL spectra of F1, F2 and F3 (650°C) a) at lower energy and b) higher energy range. Kinetics for exciton decay for (c) ReS$_2$ (450°C, F1), (e) ReS$_2$ (650°C, F2), and (g) ReS$_2$ (750°C, F3). The contour plots of TA measurements for (d) ReS$_2$ (450°C, F1), (f) ReS$_2$ (650°C, F2) and (h) ReS$_2$ (750°C, F3).*

The evolution of the excited state dynamics for the films F1-F3 is probed with the picosecond resolved TA spectroscopy, where the samples are excited with a beyond band-gap pump beam of energy 3.1 eV and the resulting excitations are explored with a probe beam with energy ranging from 1.76 to 2.6 eV. Figure 8(c), (e) and (g) depicts the spectral plots representing the change in the absorbance $m\Delta A = A_{excited} - A_{ground}$, between the excited and ground states as a function of energy. The corresponding kinetics of the excitons are presented in the contour plots in Figure 8(d), (f) and (h) respectively, where the delay time between the pump and probe beams are plotted as a function of the wavelength, with the color codes in *mΔA*, presented as a side bar. All these three samples have displayed the clear presence of both A and B excitons at ~ 1.47 and 1.52 eV respectively from the transient absorption optical data of Fig 8(c), (e) and (g). For F1 and F3, exciton A is more intense than B, whereas for F2, exciton B remains more prominent. A significant bleach cum state-filling is observed for all three systems, with the indication of F2 producing the most uniform contour plot. All these figures conveyed two extremely intriguing excited state features for the CVD films. First, with the exception of the planar sample F1, both vertical growths F2 and F3 show a significantly large lifetime of excitons, with an indication of survival even more than 917 ps. The contour plots indicate significant bleaches after a delay of more than 6900 fs. The decay time and the filling up of the trap states corresponding to both of the exciton signals are plotted in Figure 9 (a) - (d), indicating persistent excitons even after 1200 ps. Such high lifetimes of excitons can be explained from the theoretical implications of Figure 5, where the extremely localized SV-induced mid-gap states act as electron traps, imparting a longer lifetime for the excitons. Secondly, for F1 and F3, with increasing delay time, there is a blue-shift in the energy of the excitons, known as the Stark shift. The occurrence of this shift can be explained by the concomitant increase of the electron-electron repulsion with the delay time, causing an increase of the required energy to generate an electron-hole pair. Interestingly, for F2, this shift is negligible and thus the optimized sample F2 produces almost constant-energy exciton peaks with significantly long lifetimes.

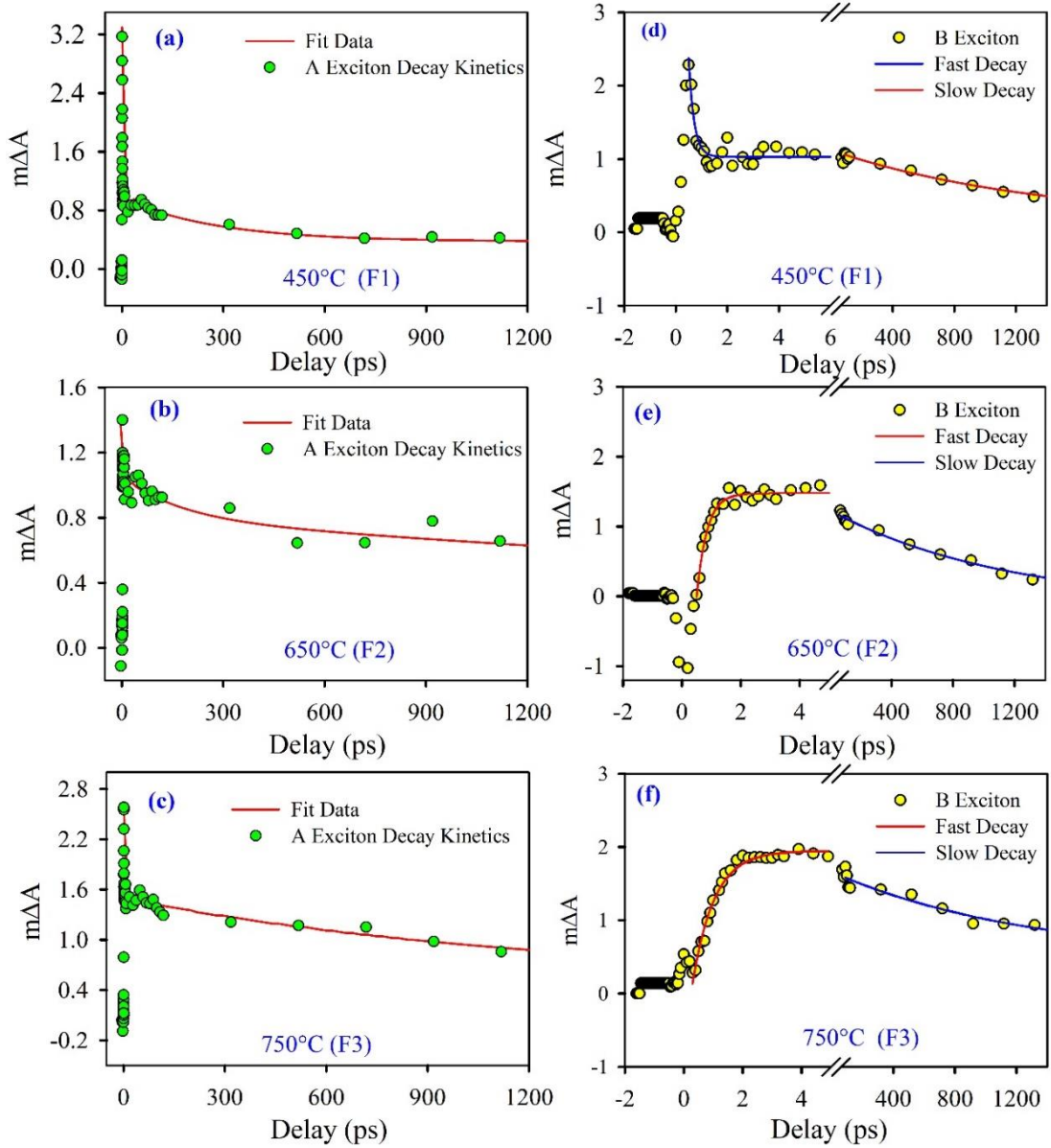

**Figure 9:** *Kinetics for A-exciton decay and trap state filling for a) F1, b) F2 and c) F3. Exciton signals are seen to be alive even after 1200 ps. Kinetics for B-exciton slow decay and fast decay and trap state filling at for d) F1, e) F2 and f) F3. Exciton signals are seen to be alive even after 1200 ps.*

In the next section, we have established the application of the vertically grown systems in the field of optoelectronic devices.

## V. Opto-electronic device application

We demonstrate the potential of vertically grown optimized system F2 in the field of optoelectronics, where, phototransistors were fabricated from F2 after using an ion-gel dielectric gate. The schematic and the corresponding characteristics of the device are depicted in Figure 10(a) – (f). For modern age integrated electronics, the ion-gel dielectrics are advantageous over the conventional solid-state dielectrics in terms of their simple fabrication protocol at room-temperature, cost-effectiveness, low operational voltage, low power consumption and the flexible choice of substrates. Figure 10 (a) presents the schematic diagram of the FET, whereas, in our practical system, the single-layer electronic material as shown in the figure is replaced by the multilayered and vertically grown system F2. Figure 10(b) depicts the transfer characteristics presenting the drain-current ($I_d$) versus gate voltage ($V_g$) curves measured at the drain-source bias ($V_d$) of 1$V$, exhibiting the typical FET behavior with distinct on-off states with a signature of *n*-type channel characteristics. The typical on-off ratio measured for such devices are ~ $10^2$ with the threshold voltage ($V_{Th}$) being ~ -8 V. The low threshold voltage and the small operating voltage window (-8 V to 3 V) for on/off switching are due to the high specific capacitance of the dielectric ion-gel, allowing an easy modulation of the work function of the $ReS_2$ channel. Details of this measurement procedure and the extracted parameters are described in the supporting information [37]. The field-effect mobility, as estimated for the $ReS_2$ channel is ~ 0.11 $cm^2V^{-1}s^{-1}$, which is comparable to the earlier reported literature [58]. For the as-grown vertically oriented films, the carrier transport is a complicated process. In the present case the active channel length of the photodetector is ~ 2 mm, which, being much larger than the nano-lithographically fabricated devices, results into a smaller value of the longitudinal electric field. Therefore, for such large-area devices, higher source-drain bias is commonly required for an efficient transport of photo-generated carriers.

The current-voltage ($I_d$ -$V_d$) characteristics as measured for fixed gate voltage ($V_g$ = - 5 V) under different illumination conditions, are presented in Figure 10(c), which indicate the symmetric I–V nature as originated from the back-to-back Schottky junction of Au and $ReS_2$ interface at the source and drain contacts. The significant rise of the current level upon illumination is attributed to the generation of photo-carriers, leading to a light-induced enhancement of the field-effect mobility of the channel. The temporal current response of the fabricated photodetector, upon periodic illuminations, has been recorded at $V_d$ = 2 V and $V_g$ = - 5 V for various illumination intensities and the results are exhibited in Figure 10(d). The rapid and periodic change in the current level for dark and illuminated conditions represents an excellent reproducibility and stability of the photodetectors. The results also demonstrated the absence of persistence photocurrent and fluctuation in the current level with time. The monotonic increase of responsivity (R) and detectivity (D) as a function of the intensity of illumination, as computed from the relation $R(\lambda) = \frac{J_{Photo}(\lambda)}{P_d}$ and $D = \frac{R(\lambda)}{\sqrt{q \cdot J_{dark}}}$ with $J_{Photo}$ and $J_{dark}$ being the photocurrent density in illuminated and dark conditions, $P_d$ is the incident power density and q is the electronic charge, are plotted in Figure 10(e). The maximum photo-responsivity was recorded as 5 A/W, while the detectivity was recorded as high as ~5×10$^{11}$ Jones, for 4.14 mW/cm$^2$ irradiation. The variation of both R and D with higher incident optical intensity becomes almost linear, suggesting that with increasing power density, the rate of photo-induced carrier generation is proportional to the absorbed photon flux. While the variation of R is linear, D undergoes saturation with respect to the bias voltage.

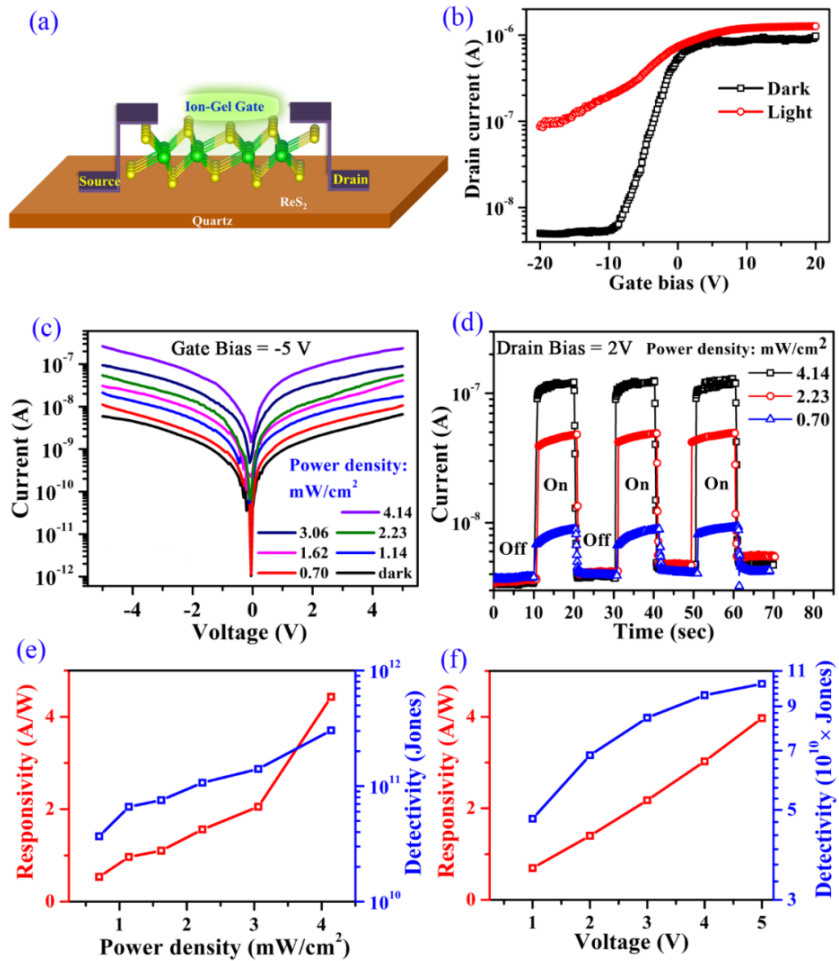

*Figure 10:* *Phototransistor characteristics of ReS$_2$ sample F2: (a) schematic device diagram, (b) transfer characteristics in dark and illuminated conditions, (c) Current versus voltage photo-response under dark and different illumination conditions, (d) Photo-transistor switching characteristics with different illuminating power, (e) responsivity and detectivity with respect to the illuminated power density of light at a gate bias of 5 V and (f) responsivity and detectivity of the system as a function of voltage.*

Therefore, our results provide an important step towards the realization of the application of the vertically grown CVD ReS$_2$ films as an efficient millimeter scale top-gate FET-based photodetector, for low-power applications in comparison to other vertically grown 2D systems [59,60].

## VI. Conclusion

With a detailed theoretical comparison of the available structures in the literature, we have identified the proper structure of ReS$_2$, capable of resolving the prior disputes about the correlation of its structure and electronic properties. Subsequently, with an appropriate bottom-up procedure, we have optimized the CVD growth conditions to obtain the sample with uniform vertical growth morphology. Microstructural and vibrational characterization identified this structure to be similar to our theoretically concluded one. The static and excited state optical properties of the films divulge a significantly long lifetime for the weakly bound excitons due to the presence of SV-induced mid-gap localization. We have demonstrated an application of these large-area vertically grown films by fabricating a photodetector with large responsivity and detectivity. The results presented in this work will be useful for the future application of vertically grown systems in the fields of optics and opto-electronics.

**Table I: Raman peak positions and EDX elemental map of different samples**

| Growth temperature (Sample Name) | Raman Peak Position | | | | | EDX of elements | |
|---|---|---|---|---|---|---|---|
| | $A_g$ (cm$^{-1}$) | $E_g$ (cm$^{-1}$) | $A_{1g}$ (cm$^{-1}$) | $E_g$-like (cm$^{-1}$) | $E_{1g}$ (cm$^{-1}$) | S (at%) | Re (at%) |
| 450°C (F1) | 149.7 | 160.0 | 210 | 304.7 | 316.8 | 64.0 | 36.0 |
| 550°C (F2) | 149.7 | 160.0 | 209.5 | 304.5 | 316.6 | 67.1 | 32.9 |
| 650°C (F3) | 150.2 | 160.26 | 210.5 | 304.4 | 316.8 | 67.5 | 32.5 |

**Table II: XPS binding energies of the Re-4$f$ and S-2$p$ electrons and the thickness of different systems**

| Samples | XPS binding energy | | | | Thickness (nm) |
|---|---|---|---|---|---|
| | Re-4$f_{5/2}$ | Re-4$f_{7/2}$ | S-2$p_{1/2}$ | S-2$p_{3/2}$ | |
| 450°C(F1) | 43.4 | 41.0 | 162.7 | 161.5 | 434 |
| 650°C(F2) | 43.0 | 40.6 | 162.3 | 161.1 | 445 |
| 750°C(F3) | 42.8 | 40.4 | 162.1 | 160.9 | 441 |
| **Ref. 44** | 44.58 | 42.18 | 162.78 | 161.74 | - |

## Acknowledgements and Author Contribution

T.K.M. acknowledges the support of DST India for the INSPIRE Research Fellowship and SNBNCBS for funding. D.K. would like to acknowledge the BARC ANUPAM supercomputing facility for computational resources, BRNS CRP on Graphene Analogues for support and motivation, SAIF-IITB for TEM instrument, Madangopal Krishnan and Abhijit Ghosh for helps in accessing the departmental facilities.

JRG, AK and RH had participated in CVD deposition, TKM and AK have plotted the figures, participated in analyzing both theoretical and experimental data, AK and KVA have done the TPA measurements and the corresponding data analysis, SM and SKR have performed the device fabrication and measurements, RA and KD have helped in Raman measurement, AMRS and AN have done the XPS and profilometry, SD and KM have measured the PL, DK has written the manuscript, performed DFT and TDDFT calculations, participated in measurements and data-analysis.

## Supporting Information

See supporting information for details of the acquired methodologies for DFT and TDDFT calculations, anisotropic optical absorbance, details of CVD process and description of device fabrication and electrical measurements, description of PL and TA measurements. It also includes additional References [61-64].

*Supplemental Material for*
# Two-dimensional ReS$_2$: Solution to the Unresolved Queries on Its Structure and Inter-layer Coupling Leading to Potential Optical Applications

Janardhan Rao Gadde[+,1], Anasuya Karmakar[+,2], Tuhin Kumar Maji[+,3], Subhrajit Mukherjee[4], Rajath Alexander[5], Anjanashree M R Sharma[6], Sarthak Das[7], Anirban Mandal[8], Kinshuk Dasgupta[5], Akshay Naik[6], Kausik Majumdar[7], Ranjit Hawaldar[1], K V Adarsh[8], Samit Kumar Ray[4,9] and Debjani Karmakar*[,10]

[1] *Centre for Materials for Electronic Technology, Pune, India, 411008*
[2] *Indian Institute of Science Education and Research, Pune, India, 411008*
[3] *Department of Chemical Biological and Macromolecular Sciences, S.N. Bose National Centre for Basic Sciences, Salt Lake, Kolkata, India, 700106*
[4] *Department of Physics, Indian Institute of Technology Kharagpur, Kharagpur, India, 721302*
[5] *Advanced Carbon Materials Section, Bhabha Atomic Research Centre, Trombay, Mumbai, India, 400085*
[6] *Centre for Nano Science and Engineering, Indian Institute of Science, Bangalore, India, 560012*
[7] *Department of Electrical Communication Engineering, Indian Institute of Science, Bangalore, India, 560012*
[8] *Department of Physics, Indian Institute of Science Education and Research, Bhopal, India 462066*
[9] *Department of Condensed Matter Physics and Material Science, S.N. Bose National Centre for Basic Sciences, Salt Lake, Kolkata, India, 700106*
[10] *Technical Physics Division, Bhabha Atomic Research Centre, Trombay, Mumbai, India, 400085*

[+] *These authors have equal contribution to this work*
*\*Corresponding Author: Dr. Debjani Karmakar ([debjan@barc.gov.in](mailto:debjan@barc.gov.in))*

**Computational Details :**

First principles calculations on different structures are performed by using the Quantumwise ATK 15.1 package[1,2], where norm-conserved FHI pseudopotentials and double-zeta polarized plane-wave basis set are used for the expansion of electronic density after incorporating the dipolar correction as per Grimme DFT-D3 method[3]. The exchange correlation energies are calculated within generalized gradient approximation (GGA) after including spin-orbital coupling and with a hybrid BLYP functional. The plane-wave cut-off energy is kept at 100 Hartree and *k*-point samplings are done over a Monkhorst-Pack mesh of 5×5×3. All the structural optimizations are done by using a LBFGS optimizer with a force and stress tolerance of 0.01 eV/A and 0.001 eV/A$^3$ respectively.

The time-dependent optical properties were calculated after including all transitions by using the all-electron full-potential linearized augmented plane wave approach, including the local orbitals (FP-LAPW + lo) within the framework of TDDFT as implemented in the ELK 4.3.6 code[4]. We treated the exchange–correlation potentials with local spin-density approximation (LSDA) with the Perdew–Wang/Ceperley Alder functional.

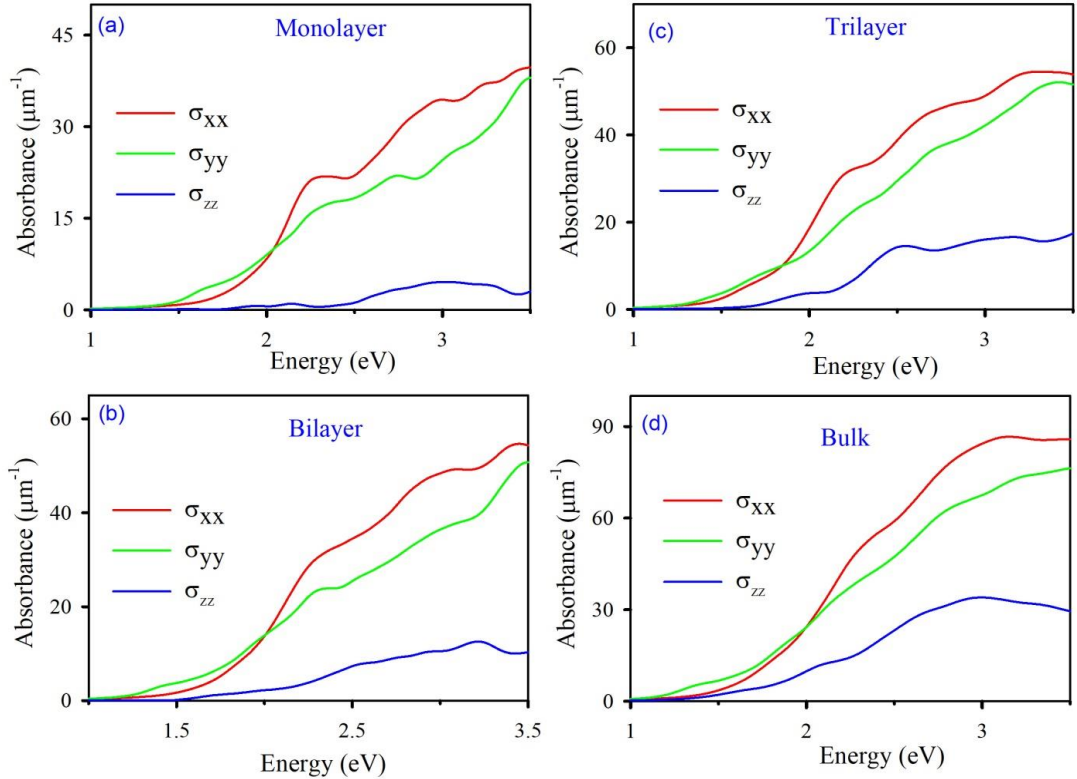

*Figure S1:* *Static DFT absorbance of different ReS$_2$ system with all three polarization dependent components for a) monolayer, b) bilayer, c) trilayer and d) bulk. Strong anisotropy present in the system is evident from the figure.*

**Chemical Vapor deposition process:**

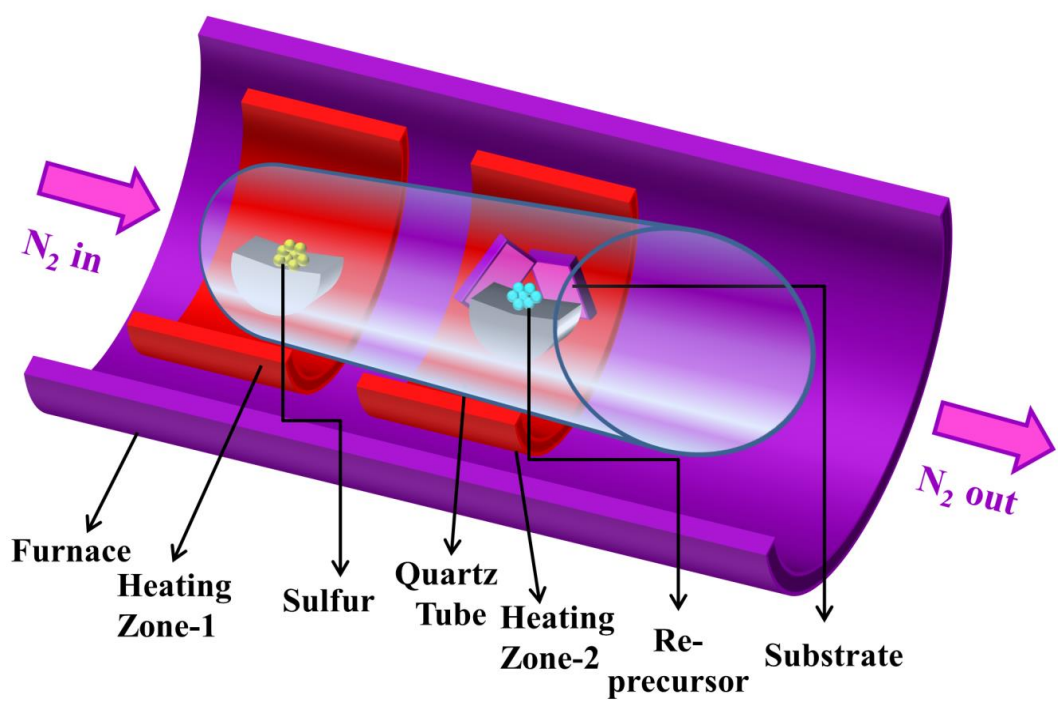

*Figure S2: The schematic for the CVD growth to produce the ReS$_2$ thin films on Si/SiO$_2$ substrates.*

In the Chemical Vapor Deposition (CVD) process to grow the ReS$_2$ thin films, the Re-precursor (NH$_4$ReO$_4$) is placed in a ceramic boat and sulfur powders (S) are placed in a glass boat. The Si/SiO$_2$ or quartz substrates were placed on top of the ceramic boat as shown. The ceramic and glass boats are then carefully placed within the CVD glass chamber, towards the center and near the edge respectively. The deposition were carried out under a continuous flow of N$_2$ gas within the sealed CVD chamber at three different temperatures of 450°C, 650°C and 750°C.

**Device fabrication and measurement:**

For modern age integrated electronics, the ion-gel dielectrics are advantageous over the conventional solid-state dielectrics in terms of their simple fabrication protocol at room-temperature, cost-effectiveness, low operational voltage, low power consumption and the flexible choice of substrates. The SiO$_2$ (300 nm) on the Si surface serves as the insulating substrate for device fabrications. Polyethylene oxide (PEO) and lithium perchlorate (LiClO$_4$) were used to prepare the gel-electrolyte as a dielectric layer for the transistors. At first, the precursors were taken in 8:1 (PEO:LiClO$_4$) weight ratio and mixed thoroughly using a magnetic stirrer at 1000 rpm. The 80 nm thick metal (Au) electrodes were deposited by thermal evaporation technique at the base pressure of ~2 x $10^{-6}$ mbar. A separate coplanar metal finger positioned remotely from the channel, was served for gate electrode, which control the electric field of the channel through ion gel. The prepared dielectric gel was carefully drop-cast in the fashion that caps the transistor channel and a portion of the gate electrode. All the electrical and photo-response measurements were performed at ambient conditions. The current-voltage (I-V) characteristics were recorded after using a Keithley 4200-SCS semiconductor parameter analyzer unit. The photocurrent was recorded using a broadband light source, a monochromator and an optical microscope equipped with an optical fibre. The low threshold voltage and the small operating voltage window (-8 V to 3 V) for on/off switching are attributed to the high specific capacitance of the dielectric ion-gel, allowing an easy modulation of the work function of the $ReS_2$ channel.

The generic fabrication process of the ion-gel dielectrics provides significant flexibility to choose any architectures or flexible substrates. After application of a positive gate bias ($V_g$) to the channel material ($ReS_2$), the resultant electric field drives $ClO_4^-$ ions towards the gate electrode and $Li^+$ ions near the channel surface, forming a nanoscale thickness gate capacitor, knows as an electric double layer (EDL). The large capacitance of the EDL leads to a large surface carrier density and induced electrons in the *n*-type channel, effectively enhancing the electric current through the channel between the source and the drain electrodes. The field-effect mobility has been extracted by using the following expression:

$$\mu = \frac{\nabla I_d}{\nabla V_g} \times \frac{L}{W C_{sp} V_d}$$

where L is the channel length (2 mm), W is the channel width (2 mm), and $C_{sp}$ is the specific capacitance of the dielectric gel, about ~3 µF/cm$^2$. The field-effect mobility was estimated as ~0.11 cm$^2$V$^{-1}$s$^{-1}$ for the present system.

**Details of PL and TA**
PL was measured with a laser of wavelength 532 nm, a spot size of 20 µm diameter and the maximum laser power being 1.6 mW. The evolution of the excited state dynamics for the films F1-F3 is probed with the picosecond resolved TA spectroscopy, where the samples are excited with a beyond band-gap pump beam of energy 3.1 eV and the resulting excitations are explored with a probe beam with energy ranging from 1.76 to 2.6 eV.